\documentclass{article}

\usepackage{arxiv}

\usepackage[utf8]{inputenc} 
\usepackage[T1]{fontenc}    
\usepackage{hyperref}       
\usepackage{url}            
\usepackage{booktabs}       
\usepackage{amsfonts}       
\usepackage{nicefrac}       
\usepackage{microtype}      
\usepackage{lipsum}
\usepackage{graphicx}
\usepackage{epsfig}
\usepackage{subcaption}
\usepackage{xcolor}
\usepackage{pifont}
\usepackage{amsmath,amssymb} 
\usepackage{multirow}
\usepackage{upgreek}
\usepackage{arydshln}
\usepackage{tikz}
\usepackage{comment}
\usepackage{color}
\DeclareMathAlphabet{\pazocal}{OMS}{zplm}{m}{n}

\title{Extreme Generative Image Compression by Learning Text Embedding from Diffusion Models}

\author{
  Zhihong Pan, Xin Zhou, Hao Tian \\
  Baidu Research (USA)\\
   \\
}

\begin{document}
\newcommand{\cmark}{\ding{51}}%
\newcommand{\xmark}{\ding{55}}%
\newcommand{\red}[1]{\textcolor{red}{#1}}
\newcommand{\blue}[1]{\textcolor{blue}{#1}}
\newcommand{\green}[1]{\textcolor{green}{#1}}
\newcommand{\teal}[1]{\textcolor{teal}{#1}}
\newcommand{\orange}[1]{\textcolor{orange}{#1}}
\newcommand{\wt}[1]{\textcolor{white}{#1}}
\newcommand{\etal}{\textit{et al.}}

\maketitle

\begin{abstract}
Transferring large amount of high resolution images over limited bandwidth is an important but very challenging task.  Compressing images using extremely low bitrates (\textless0.1 bpp) has been studied but it often results in low quality images of heavy artifacts due to the strong constraint in the number of bits available for the compressed data.  It is often said that a picture is worth a thousand words but on the other hand, language is very powerful in capturing the essence of an image using short descriptions.  With the recent success of diffusion models for text-to-image generation, we propose a generative image compression method that demonstrates the potential of saving an image as a short text embedding which in turn can be used to generate high-fidelity images which is equivalent to the original one perceptually.  For a given image, its corresponding text embedding is learned using the same optimization process as the text-to-image diffusion model itself, using a learnable text embedding as input after bypassing the original transformer.  The optimization is applied together with a learning compression model to achieve extreme compression of low bitrates \textless0.1 bpp.  Based on our experiments measured by a comprehensive set of image quality metrics, our method outperforms the other state-of-the-art deep learning methods in terms of both perceptual quality and diversity.
\end{abstract}

\section{Introduction}
\label{sec:intro}

With the increasing amount of image streams available for broad range of applications, lossy image compression is a very
useful technique for efficient image storage and transmission.  Over the years, various engineered codes such as JPEG~\cite{leger_oe_1991}, 
JPEG2000~\cite{skodras_sp_2001}, and the more recent BPG\cite{bellard_bpg} have been proposed to compress single images but their performance
have saturated overall.  More recently, deep learning based image compression methods have been
studied~\cite{balle_arxiv_2018, minnen_nips_2018, cheng_cvpr_2020}.  These models are generally trained in an end-to-end
fashion to minimize a rate-distortion object $R+\lambda D$.  Here $R$ represents the entropy of latent representations which
is estimated by an entropy model, $D$ is the difference between the original image and the compressed one, and $\lambda$
determines the desired trade-off between rate and distortion.  When $\lambda$ is small, the optimization gives higher priority
to compression rate so the resulted bitrate (evaluated as bits-per-pixel, bpp) is low.  Consequently, the compressed image has
lower quality due to higher $D$ loss term.  With accuracy metrics like mean squared error (MSE) and multi-scale structural
similarity (MS-SSIM) are often used for $D$, the low quality compressed images are usually blurry. For extremely low
bitrates (\textless0.1 bpp), both engineered codecs and deep learning compression models are subject to very poor perceptual qualities.

To tackle this problem, some recent methods~\cite{tschannen_nips_2018, wu_wacv_2020, lee_cvprw_2020, mentzer_nips_2020}
aim to restore less blurry image from highly compressed latent representations at the cost of accuracy.
These model adopt generative adversarial networks (GAN)~\cite{goodfellow_nips_2014} to fully or partially replace the accuracy metrics in
$D$ with discrimination loss so they can generate sharp and realistic images even at very low bitrates.
For the challenging task of extremely low bitrates, GAN is further exploited
in more recent studies~\cite{agustsson_iccv_2019, dash_wacv_2020, iwai_icpr_2021} to restore sharp images with minimized distortion
and visual artifacts.  However, they all inherit the drawback of unstable training from GAN, making it difficult to tune the training process for large datasets. In this paper, we propose the first generative image compression
method with extremely low bitrates using denoising diffusion models.  As it utilizes an existing text-to-image model which is
already trained with a gigantic dataset, it is applicable to any type of image with no need of further tuning.

\begin{figure*}[t]
\captionsetup[subfigure]{font=small, labelformat=empty}
\begin{center}
  \begin{subfigure}[b]{0.12\textwidth}
    \centering
      \includegraphics[width=\textwidth, interpolate=false]{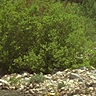}
  \end{subfigure} \hspace*{-0.5em}
  \begin{subfigure}[b]{0.12\textwidth}
    \centering
      \includegraphics[width=\textwidth, interpolate=false]{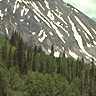}
  \end{subfigure} \hspace*{-0.5em}
  \begin{subfigure}[b]{0.12\textwidth}
    \centering
      \includegraphics[width=\textwidth, interpolate=false]{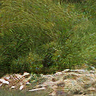}
  \end{subfigure} \hspace*{-0.5em}
  \begin{subfigure}[b]{0.12\textwidth}
    \centering
      \includegraphics[width=\textwidth, interpolate=false]{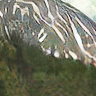}
  \end{subfigure} \hspace*{-0.5em}
  \begin{subfigure}[b]{0.12\textwidth}
    \centering
      \includegraphics[width=\textwidth, interpolate=false]{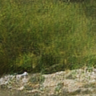}
  \end{subfigure} \hspace*{-0.5em}
  \begin{subfigure}[b]{0.12\textwidth}
    \centering
      \includegraphics[width=\textwidth, interpolate=false]{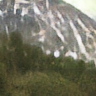}
  \end{subfigure} \hspace*{-0.5em}
  \begin{subfigure}[b]{0.12\textwidth}
    \centering
      \includegraphics[width=\textwidth, interpolate=false]{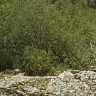}
  \end{subfigure} \hspace*{-0.5em}
  \begin{subfigure}[b]{0.12\textwidth}
    \centering
      \includegraphics[width=\textwidth, interpolate=false]{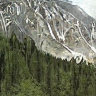}
  \end{subfigure}
  
  \begin{subfigure}[b]{0.24\textwidth}
    \centering
      \includegraphics[width=\textwidth, interpolate=false]{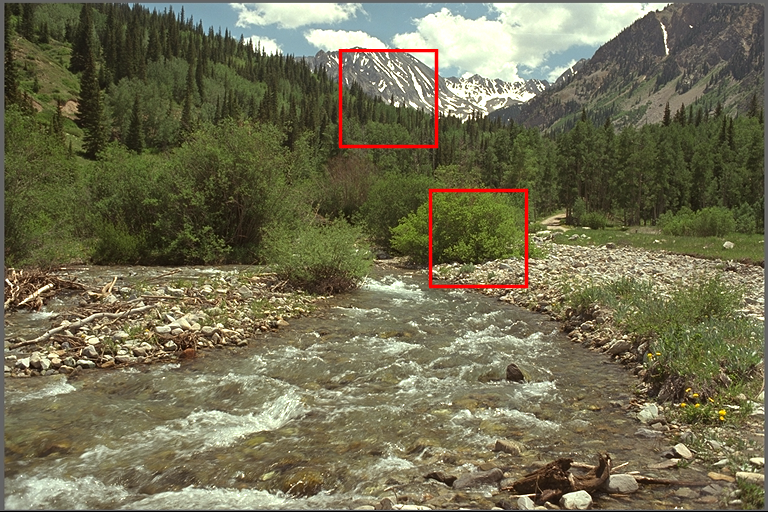}
      \caption{Original}
  \end{subfigure} \hspace*{-0.5em}
  \begin{subfigure}[b]{0.24\textwidth}
    \centering
      \includegraphics[width=\textwidth, interpolate=false]{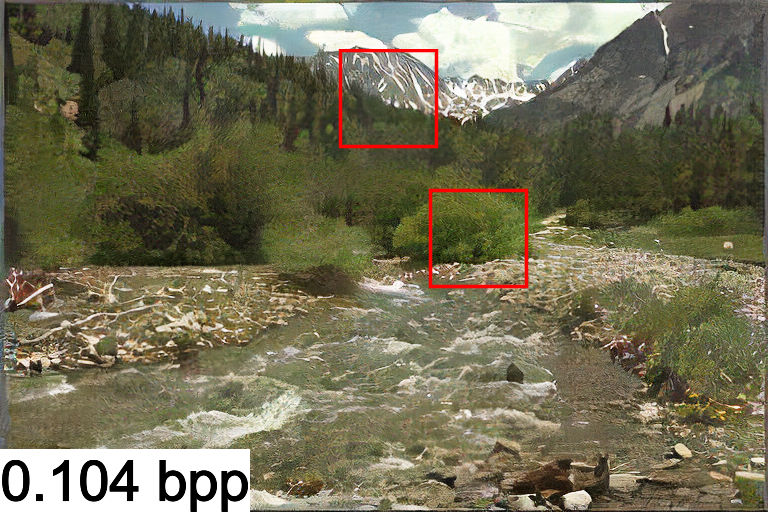}
      \caption{Iwai \etal~\cite{iwai_icpr_2021}}
  \end{subfigure} \hspace*{-0.5em}
  \begin{subfigure}[b]{0.24\textwidth}
    \centering
      \includegraphics[width=\textwidth, interpolate=false]{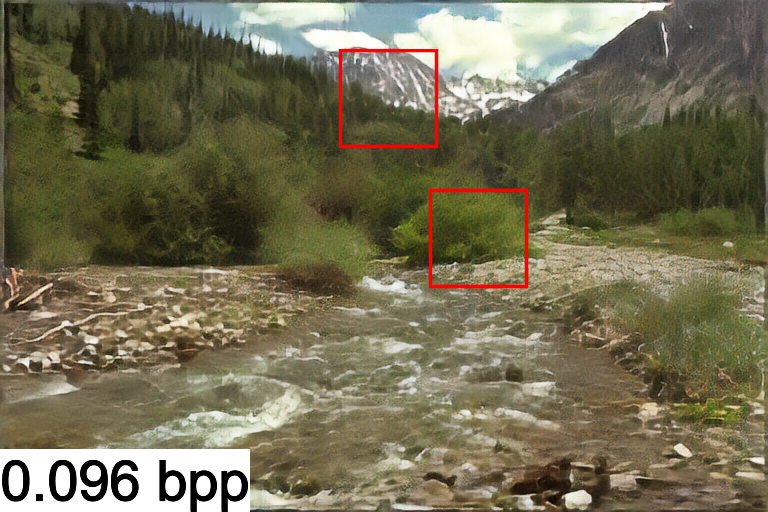}
      \caption{Mentzer \etal~\cite{mentzer_nips_2020}}
  \end{subfigure} \hspace*{-0.5em}
  \begin{subfigure}[b]{0.24\textwidth}
    \centering
      \includegraphics[width=\textwidth, interpolate=false]{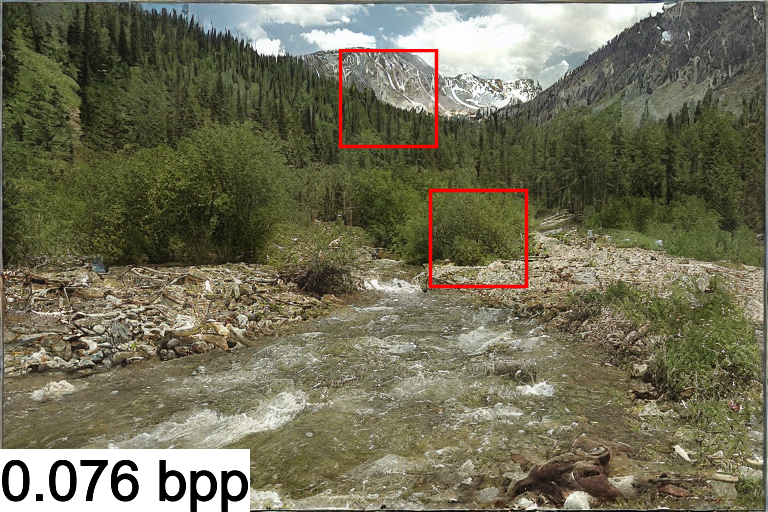}
      \caption{Ours}
  \end{subfigure}
 \end{center}
   \vspace{-9pt}
    \caption{Visual examples of generated images after compression with extremely low bitrates, demonstrating our method's superior capability to generate very sharp details.}
   \vspace{-6pt}
\label{fig:first}
\end{figure*}

Similar to GAN a few years back, denoising diffusion models~\cite{sohl_icml_2015, ho_nips_2020, song_iclr_2021} are
gaining popularity increasingly for their advantages in generating images
with high qualities in both fidelity and diversity without disadvantage of unstable training like GAN.
In addition to unconditional image generation, diffusion models have also empowered the breakthrough developments
in diffusion-based text-to-image
generation models~\cite{rombach_cvpr_2022, nichol_arxiv_2022, ramesh_arxiv_2022, saharia_arxiv_2022}
which are able to create realistic images according to given text descriptions.
They often use existing transformer models to encode text prompts as textual embeddings and use
them as conditions in training and sampling of the diffusion model.
Realising the power of these model in turning short texts into high resolution and high quality images, the generative image compression
we propose here encodes an input image as a textual embedding which is quantized and compressed for storage.  At inference time
for decompression, the compressed textual embedding is decoded and used as the conditional input for 
image generation.  While sampling from the diffusion model can generate high quality images, the random sampling could lead to
diverse outcomes without guarantee of resembling the original input.  To ensure the generation success,
the original image is also compressed from an existing learning based compression model
and it is used as a guidance at sampling time, on top of the classifier-free guidance.  The additional bitrates for compression
guidance is only $0.01-0.02$ bpp so the overall bitrates is still very low.
Note that there are a couple of newly proposed methods~\cite{theis_arxiv_2022, yang_arxiv_2022}
which are concurrently working on lossy image compression using diffusion models.
However, not like our method here, they both need ground-up training of a dedicated diffusion model and do not address the challenge of
extremely low bitrates.  Based on our experiments, our proposed model is capable of generating image of the highest perceptual
quality while maintaining overall resemblance with the original image.  As shown in Fig.~\ref{fig:first},
the other state-of-the-art methods~\cite{iwai_icpr_2021, mentzer_nips_2020} are subject to blurry artifacts when the
bitrate gets below 0.1 while our method is able to generate very sharp details.  In the example of snowy mountains,
our generated sample has details sharper even than the original.  Although details like the number of snowy tracks are different,
which results in poor measurements in terms of pixel by pixel accuracy, our sample is highly photo-realistic and look like the same
image as the original overall.

In summary, we propose an innovative generative image compression method with the following main contributions:
\begin{itemize}

\setlength\itemsep{0.01em}
\item[$\bullet$] Using existing text-to-image diffusion models, our method can compress an input image as a textual embedding 
of extremely low bitrates and later generate diverse diverse sharp images which resemble the input perceptually.

\item[$\bullet$] A hybrid guidance method is studied to combine classifier-free guidance from pre-trained text-to-image
models and newly introduced compression guidance for optimal generation results.

\item[$\bullet$] The number of bits needed to compress the textual embedding is largely independent of image content and resolution, so
the bitrate is relatively constant for a fixed resolution and decrease when the image resolution increases.

\end{itemize}

\section{Related Works}
\label{sec:rwork}

\subsection{Image Compression}
Shannon’s theory of communication~\cite{shannon_1948} has provided the fundamental basis for the coding theory used
in classical image compression methods.   Using explicit probabilistic modeling and feature extractions,
various codes, like JPEG~\cite{leger_oe_1991}, BPG~\cite{bellard_bpg} and WebP~\cite{google_webp},
have been effectively engineered for the task of image compression.
The earliest learning based image compression methods~\cite{toderici_arxiv_2015, toderici_cvpr_2017} relied on RNNs~\cite{medsker_2001}.
Ball{\'e} \etal~\cite{balle_arxiv_2018} were the first to introduce
an end-to-end autoencoder and entropy model that jointly optimizes rate and distortion,
which was then enhanced with a scale hyperprior in \cite{balle_arxiv_2018} to
capture spatial dependencies in the latent representation.
Later various autoregressive and hierarchical priors~\cite{minnen_nips_2018, hu_aaai_2020}
were introduced to further improve the compression performance.
Cheng \etal~\cite{cheng_cvpr_2020} added attention modules and used a Gaussian Mixture Model
(GMM) to estimate the latent representation distribution for further improvements.

 \begin{figure*}[t]
 \begin{center}
     \includegraphics[width=\linewidth]{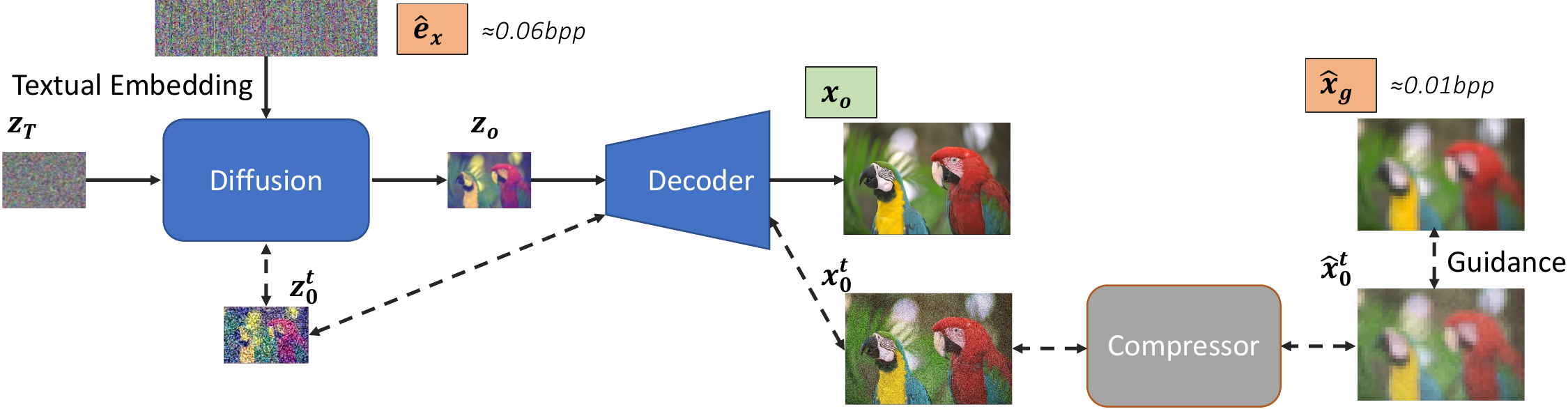}
 \end{center}
 \vspace{-3pt}
 \caption{Overview of the sampling process of proposed generative image compression using two inputs of extremely low bitrates, where $\hat{e}_x$, a highly compressed textual embedding, is used as the conditional input for a pre-trained latent diffusion model, and $\hat{x}_g$, a highly compressed image from original image $x$, is used a constraint to guide the intermediate latent image $z^t_0$ at each time step $t$.  These two are saved after the initial compression process and are the only two needed to reconstruct a high quality image $x_0$.}
 \vspace{-12pt}
 \label{fig:process}
 \end{figure*}

Since their introduction in~\cite{goodfellow_nips_2014},
GANs have progressed greatly in unconditional and conditional image generation
of high resolution photo-realistic images~\cite{donahue_arxiv_2018, tulyakov_cvpr_2018, karras_cvpr_2019, karras_nips_2021}.
The adversarial loss function was first introduced in an end-to-end framework~\cite{rippel_icml_2017} for
improved perceptual quality, and has been continuously improved in following
studies~\cite{santurkar_pcs_2018, tschannen_nips_2018}.
While these methods are capable of reconstructing photo-realistic image with very low bitrate,
generative image compression with extremely low bitrates (\textless0.1 bpp) was first studied
in~\cite{agustsson_iccv_2019} and further improved in following studies~\cite{dash_wacv_2020, iwai_icpr_2021}.
Comparing to these GAN based extreme generative compression models, ours is the first to utilize diffusion
models to tackle this challenging task.

\subsection{Denoising Diffusion Models}
Inspired by non-equilibrium thermodynamics~\cite{sohl_icml_2015}, the denoise diffusion models
define a Markov chain of diffusion steps to slowly add random noise to data so the intractable real data distribution
is transformed to a tractable one like Gaussian.  Then the models learn to reverse the diffusion process to construct desired data samples from
randomly sample Gaussian noise.  Ho \etal~\cite{ho_nips_2020} proposed a denoising diffusion probabilistic model (DDPM) to interpret
the reverse diffusion process as a large amount of consecutive denoising steps following conditional Gaussian distribution.
Alternatively, Song \etal~\cite{song_nips_2019, song_iclr_2020} used stochastic differential equations to model the reverse diffusion process
and developed a score-based generative model to produce samples via Langevin dynamics using estimated gradients of the data distribution.
Later numerous methods~\cite{nichol_icml_2021, song_iclr_2021, lu_arxiv_2022} have been proposed to use much fewer denoising steps
without significant degradation in image quality.  To improve image quality,
Dhariwal \etal~\cite{dhariwal_nips_2021} proposed a classifier guidance method to iteratively modify the denoised step using a gradient
calculated from a retrained noisy classifier.
Later Ho \etal~\cite{ho_nipsw_2021} invented a classifier-free guidance method
that trains a conditional model using randomly masked class labels and treat the difference between conditional and unconditional sampling at
inference time as a proxy classifier.  The compression guidance proposed here is applied similarly as the classifier guidance.

In recent years, GAN based deep learning models have been successful used for various
generative tasks~\cite{donahue_arxiv_2018, tulyakov_cvpr_2018, karras_cvpr_2019},
including text-to-image generations~\cite{reed_icml_2016, zhang_iccv_2017, xu_cvpr_2018, qiao_cvpr_2019, tao_arxiv_2020, frolov_nn_2021}.
More recently, autoregressive (AR) models
have also shown promising results in image generation~\cite{parmar_icml_2018, chen_icml_2020, esser_cvpr_2021}.  For text-to-image
generations, various frameworks, including DALL-E~\cite{ramesh_icml_2021}, CogView~\cite{ding_arxiv_2021} and M6~\cite{lin_arxiv_2021},
have been proposed to use large transformer structure to model the joint distribution of text and image tokens.
Diffusion models have progressed rapidly to set state-of-the-art for many generative tasks, including text-to-image generations.
Previously, text-to-image generation are dominated by
GANs~\cite{reed_icml_2016, zhang_iccv_2017, xu_cvpr_2018, qiao_cvpr_2019, tao_arxiv_2020, frolov_nn_2021}
and autoregressive (AR) models~\cite{ramesh_icml_2021, ding_arxiv_2021, lin_arxiv_2021}.
Most recently, diffusion-based text-to-image generation has been a red hot research topic in both the academia and industry.
Initially, an unconditional diffusion model~\cite{dd_github_2022} was demonstrated
highly capable of text-to-image generation using sampling guidance to match the CLIP scores\cite{radford_icml_2021}
of the text input and generated image.  More recent models all use
transformer based text embedding to train the conditional diffusion model,
generating either a low-resolution image~\cite{nichol_arxiv_2022, saharia_arxiv_2022} or
an image embedding \cite{saharia_arxiv_2022} before generating the full resolution output.
Alternatively, Rombach \etal~\cite{rombach_cvpr_2022} proposed to conduct the conditional text-to-image diffusion
in a latent space of reduced resolution for faster training and sampling.  Based on that, a large
text-to-image model, Stable Diffusion~\cite{sd_github_2022}, is trained with a huge dataset and released for open
research.  Our proposed image compression method is validated using the released version
$\texttt{v1-4}$.

\section{Proposed Method}
\label{sec:method}

 Our proposed generative compression method is built on pre-trained text-to-image diffusion models without any change
 in their model weights.  For demonstrative purposes, the publicly available Stable Diffusion model~\cite{sd_github_2022}
 which uses the latent diffusion model~\cite{rombach_cvpr_2022} architecture, is adopted for all experiments.
 In this case, diffusion process is conducted in the latent space $z$ which can be decoded to the image space.
 For any input image $x$, it is first applied with a state-of-art learned compressor~\cite{cheng_cvpr_2020}
 to save as a highly compressed image $\hat{x}_g$ which is used as a guidance during inference time explained later.
 It only consumes storage of a tiny portion of the original size, around 0.01 bpp.  Additionally, adopting the
 textual inversion used in prior works~\cite{gal_arxiv_2022, ruiz_arxiv_2022, kawar_arxiv_2022},
 a textual embedding $e_x$ is found as the optimal conditional input to generate $x$ from random
 noise $z_T$ using an iterative learning process.  $e_x$ is then further quantized
 and compressed as $\hat{e}_x$ using a compressor~\cite{cheng_cvpr_2020}.  As the textual embedding has
 a fixed dimensionality in regardless of the image size, the bitrate needed to save $\hat{e}_x$ depends
 on the original image size and it decreases when the number of pixels increases.  For a $512\times 768$
 image like in the Kodak dataset~\cite{franzen_kodak}, the bitrate of  $\hat{e}_x$ is around 0.06 bpp.  For large images in need of
 extreme compression, the total bitrate is expected to be 0.07 or less.
 For the generation process of decompression, as shown in Fig.~\ref{fig:process}, the denoising step in latent
 space to transform random noise $z_T$ to a latent sample $z_0$ is conditional to the compressed embedding $\hat{e}_x$.
 At the same time, for each denoising step $t$, the noise estimation is guided by correcting the
 intermediate latent sample $z^t_0$ in reference to the compressed guidance $\hat{x}_g$.  After generation
 of the latent sample $z_0$, an output image $x_0$ is reconstructed through the decoder in the latent diffusion model.
 Details of the textual inversion and compression guidance are included below.
 
 \begin{figure*}[t]
 \begin{center}
     \includegraphics[width=\linewidth]{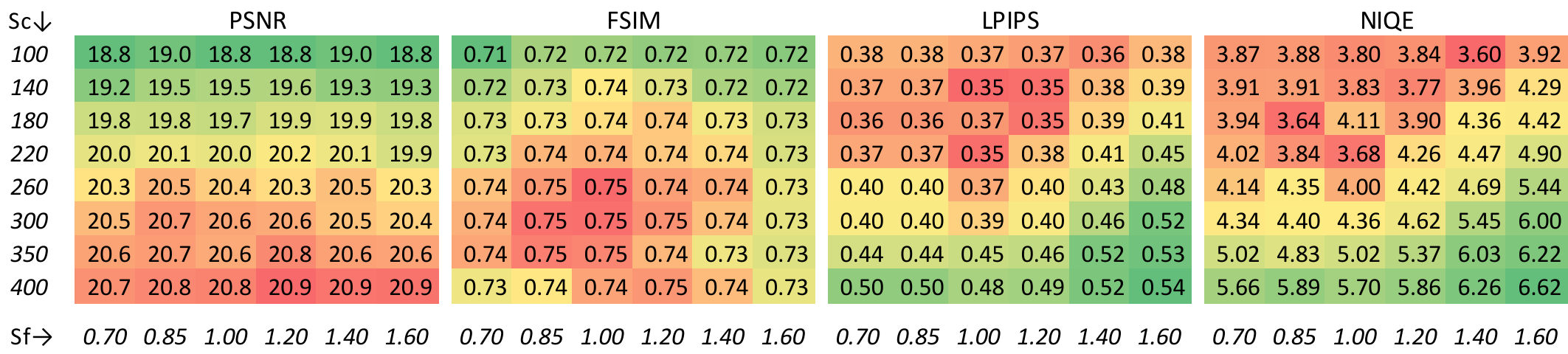}
 \end{center}
 \vspace{-12pt}
 \caption{Comparison of multiple image quality metrics for different combinations of compression guidance scale ($s_c$) and classifier-free guidance scale ($s_f$).  Red in the heatmaps means higher quality while green meas lower.}
 \vspace{-6pt}
 \label{fig:scales}
 \end{figure*}

\subsection{Latent Diffusion Model Background}
For any image $x_0$, it is first encoded as a latent sample $z_0$ and the distribution of $z_0$ from all realistic images is denoted
as $q(z_0)$.  Each $z_0$ can be transformed progressively as $z_1, ..., z_T$ through added Gaussian noises and when $T$ is large enough $z_T$ becomes a random noise.
In order to convert any random noise $z_T$ to a latent sample $z_0$, 
a diffusion model with parameter $\theta$ is designed to match the true posterior using:
 \begin{equation}
 \resizebox{\width}{!}{$p_{\theta}(z_{t-1}|z_t) = \mathcal{N}(\mu_{\theta}(z_t, t), \Sigma_{\theta}(z_t, t))$}.
 \end{equation}
Here $\Sigma_{\theta}(z_t, t)$ is often fixed as $\sigma_t \mathrm{I}$ in practice so the diffusion model is only trained
for $\mu_{\theta}(z_t, t)$.  
Starting from a noise $z_T \sim \mathcal{N}(0, \mathrm{I})$, the learned posterior can
be used to sample $z_t$, $t=z-1, z-2, ...$ iteratively.  The output latent sample latent $z_0$
can be decoded as a generated image $x_0$.

As shown in DDPM~\cite{ho_nips_2020}, a re-weighted variational lower-bound (VLB) is used as an effective surrogate objective for diffusion model optimization.
As $\mu_{\theta}(z_t, t)$ can be derived from $\epsilon_{\theta}(z_t, t)$, estimated
noise added in $z_t$, a diffusion model can be be optimized using a simple standard mean-squared error (MSE) loss
 \begin{equation}
 \resizebox{\width}{!}{$L^{\text{simple}} = E_{t, z_0, \epsilon} ||\epsilon - \epsilon_{\theta}(z_t, t)||^2$}
 \label{eq:ls}
 \end{equation}
where $\epsilon$ is the known Gaussian noise added to $z_0$
to synthesize $z_t$.

\subsection{Compression Guidance}
In order to improve image sampling quality at inference time, Dhariwal \etal~\cite{dhariwal_nips_2021} proposed
a classifier guidance method to perturb the estimated mean
 by adding the gradient of the log-probability~$\log p_{\phi}(y|z_t)$ of a target class~$y$ predicted by a classifier where $z_t$ needs to be decoded as an image $x_t$, denoted as $x_t=d(z_t)$ where $d$ stands for
 the decoding process, before feeding to classifier in the case of
 of latent diffusion.
The resulting new perturbed mean~$\hat{\mu}_{\theta}(z_t, t|y)$ is given by
 \begin{equation}
 \resizebox{\width}{!}{$\hat{\mu}_{\theta}(z_t, t|y) = \mu_{\theta}(z_t,t|y) + s \Sigma_{\theta}(z_t,t|y) \nabla_{z_t} \log p_{\phi}(y|z_t)$}
 \end{equation}
where coefficient $s$ is called the guidance scale.  A larger $s$ leads to higher sample quality but less diversity. $\phi$ represents the classifier parameters
which can be further refined with noisy images and conditional to $t$ as $x_t$ is normally noisy.

For our proposed compression guidance, the reference to use
in place of label $y$ is an extremely low bitrate image $\hat{x}_g$, which is compress from $x$ as $\hat{x}_g = c(x)$.  Similar to classifier guidance, 
the estimated mean during the reverse denoising process is perturbed by the gradient of the difference between
$\hat{x}_g$ and compressed $x_t$:
 \begin{equation}
 \resizebox{\width}{!}{$\hat{\mu}_{\theta}(z_t, t) = \mu_{\theta}(z_t,t) - s \Sigma_{\theta}(z_t,t) \nabla_{z_t} \left|\hat{x}_g-\hat{x}_t \right|$}
 \end{equation}
where $\hat{x}_t=d(c(z_t))$.  However, unlike classifier
guidance where $\phi$ can be optimized for noisy images,
there is no learnable variable to optimize in the case of
our compression guidance.  To mitigate the impact of
noise present in $\hat{x}_t$, 
here we propose an alternative guidance method to calculate
the perturbing gradient by comparing the "noise-free" $\hat{x}^t_0$ and reference $\hat{x}_g$ instead
 \begin{equation}
 \resizebox{\width}{!}{$
 \begin{split}
 x^t_0 & = c(d((z_t-\sqrt{1-\bar{\alpha}_t}\epsilon_{\theta}(z_t, t)) / \sqrt{\bar{\alpha}_t})) \\
 \hat{\mu}_{\theta}(z_t, t) & = \mu_{\theta}(z_t,t) - s_c \Sigma_{\theta}(z_t,t) \nabla_{z_t} \left|\hat{x}_g-\hat{x}_t \right|
 \end{split}
 $}
 \label{eq:cg}
 \end{equation}
 where $\alpha_t$ is set from pre-determined noise schedule,
and $s_c$ is the compression guidance scale, to differentiate from the classifier-free guidance~\cite{ho_nipsw_2021} which is adopted in 
the diffusion model used for our experiments.  The
scale used in classifier-free guidance is denoted as $s_f$.
As both guidance methods are used together, we empirically studied the optimal settings for both $s_c$ and $s_f$ for
best effects of this hybrid guidance.

\subsection{Textual Inversion}

For the adopted textual inversion, the goal is to find an optimal textual embedding $e_x$ backwards from a given image $x$.
This process is a learning process to to minimize the following expected error:
 \begin{equation}
 \resizebox{\width}{!}{$E_{t, \epsilon} ||\epsilon - \epsilon_{\theta}(z_t, t, e_x)||^2$}
 \label{eq:le}
 \end{equation}
where $\epsilon_\theta$ is the diffusion model pre-trained with the loss term defined in Equation~\ref{eq:ls}.
with fixed weights $\theta$ and $x$ is fixed too.  By optimizing iteratively using varying $\epsilon$ and $t$, the target textual embedding $e_x$ can be learned effectively.
For any image x, the embedding $e_x$ has a fixed number of $T$ tokens and each token is embedded as a $N$-dimensional vector.  To effectively compress the $T\times N$ real numbers to meet needs of our extreme compression
application, $e_x$ is quantized and compressed with an
existing compression model~\cite{cheng_cvpr_2020},
denoted as $\hat{e}_x$.  To further optimize the whole
process, the quantization and compression process are included in the learning process of textual inversion by minimizing the following error instead
 \begin{equation}
 \resizebox{\width}{!}{$E_{t, \epsilon} ||\epsilon - \epsilon_{\theta}(z_t, t, \hat{e}_x)||^2$}.
 \label{eq:le}
 \end{equation}

\section{Experiments}
\label{sec:exp}
All experiments in this study are conducted using pre-trained Stable Diffusion~\cite{sd_github_2022} model.
For $\hat{x}_g$ used in compression guidance, the original image $x$ is first downsampled with a scale of $\times 4$
before compressed using an existing compression model~\cite{cheng_cvpr_2020} with GMM and attention.
For textual inversion, we use 64 embedded vectors, each has 768 elements.  For compression of
said textual embedding using the same existing compression model~\cite{cheng_cvpr_2020},
it is reshaped as a RGB color image of $64\times 256$ pixels.

To assess the effectiveness of our proposed method, we use the photo-realistic images from the Kodak dataset~\cite{franzen_kodak}
to conduct the compression experiments.  For each image, the optimal compressed textual embedding is determined
using 4000 iterative learning steps before tested for image generation.  To generate high
quality image at inference time, we use 100 DDIM~\cite{song_iclr_2021} sampling
steps ($\eta = 1$) for the diffusion model.
To assess the image quality quantitatively,
a comprehensive set of metrics are used.  The first set of metrics use the original image as the ground-truth (GT)
reference.  In addition to the standard PSNR metric, FSIM~\cite{zhang_tip_2011} is chosen as
a measure relying on low-level features the human visual systems often use.
and LPIPS~\cite{zhang_cvpr_2018} is also used for its effectiveness as a perceptual metric.
For blind metrics without GT reference, NIQE~\cite{mittal_spl_2012} is the one comparing image statistics
with those of undistorted images.  FID~\cite{heusel_nips_2017} and KID~\cite{binkowski_arxiv_2018} are also
no-reference metrics, calculated from statistics of learned features.
The are both chosen for their popular application on generative models and KID is known to be more robust as
an unbiased one.

\begin{table*}[t!]
	\centering
	\scriptsize
	\setlength{\tabcolsep}{4pt}
\vspace{0pt}
	\caption{Quantitative image quality comparison of  generative compression methods using Kodak dataset (best of three marked in \red{red}).}
\vspace{-3pt}
	\begin{tabular}{rccccccc} 
		\hline
		{$\wt{^{^A}}$} & {bpp} & {PSNR$\uparrow$} & {FSIM$\uparrow$} & {LPIPS$\downarrow$} & {NIQE$\downarrow$} & {FID$\downarrow$} & {KID$\downarrow$}\\
		\hline \hline
		Original$\wt{^{^A}}$ & {-} & {-} & {-} & {-} & 3.020 & 258.5 & 3.157 \\
		Iwai \etal~\cite{iwai_icpr_2021}$\wt{^{^A}}$ & 0.063 $\pm$ 0.028 & 24.39 & 0.8794 & 0.3054 & 3.566 & 269.1 & 5.379 \\
		Mentzer \etal~\cite{mentzer_nips_2020}$\wt{^{^A}}$ & 0.075 $\pm$ 0.021 & \red{24.93} & \red{0.8906} & \red{0.2037} & \red{3.563} & 264.2 & 5.471 \\
		Ours$\wt{^{^A}}$ & 0.070 $\pm$ 0.008 & 20.61 & 0.7486 & 0.3611 & 3.731 & \red{258.6} & \red{3.960} \\
		\hline	\end{tabular}
\label{tab:iqa}
\vspace{3pt}
\end{table*}

\subsection{Hybrid Guidance}

\begin{figure*}[t]
\captionsetup[subfigure]{font=small, labelformat=empty}
\begin{center}
  \begin{subfigure}[b]{0.12\textwidth}
    \centering
      \includegraphics[width=\textwidth, interpolate=false]{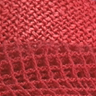}
  \end{subfigure} \hspace*{-0.5em}
  \begin{subfigure}[b]{0.12\textwidth}
    \centering
      \includegraphics[width=\textwidth, interpolate=false]{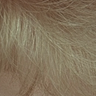}
  \end{subfigure} \hspace*{-0.5em}
  \begin{subfigure}[b]{0.12\textwidth}
    \centering
      \includegraphics[width=\textwidth, interpolate=false]{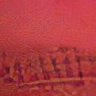}
  \end{subfigure} \hspace*{-0.5em}
  \begin{subfigure}[b]{0.12\textwidth}
    \centering
      \includegraphics[width=\textwidth, interpolate=false]{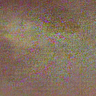}
  \end{subfigure} \hspace*{-0.5em}
  \begin{subfigure}[b]{0.12\textwidth}
    \centering
      \includegraphics[width=\textwidth, interpolate=false]{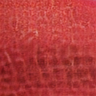}
  \end{subfigure} \hspace*{-0.5em}
  \begin{subfigure}[b]{0.12\textwidth}
    \centering
      \includegraphics[width=\textwidth, interpolate=false]{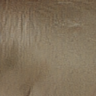}
  \end{subfigure} \hspace*{-0.5em}
  \begin{subfigure}[b]{0.12\textwidth}
    \centering
      \includegraphics[width=\textwidth, interpolate=false]{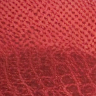}
  \end{subfigure} \hspace*{-0.5em}
  \begin{subfigure}[b]{0.12\textwidth}
    \centering
      \includegraphics[width=\textwidth, interpolate=false]{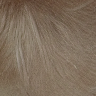}
  \end{subfigure}
  
  \begin{subfigure}[b]{0.24\textwidth}
    \centering
      \includegraphics[width=\textwidth, interpolate=false]{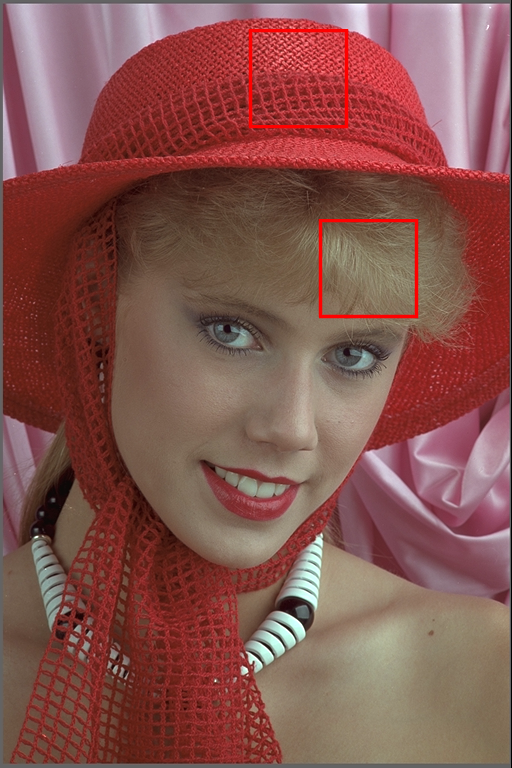}
      \caption{Original}
  \end{subfigure} \hspace*{-0.5em}
  \begin{subfigure}[b]{0.24\textwidth}
    \centering
      \includegraphics[width=\textwidth, interpolate=false]{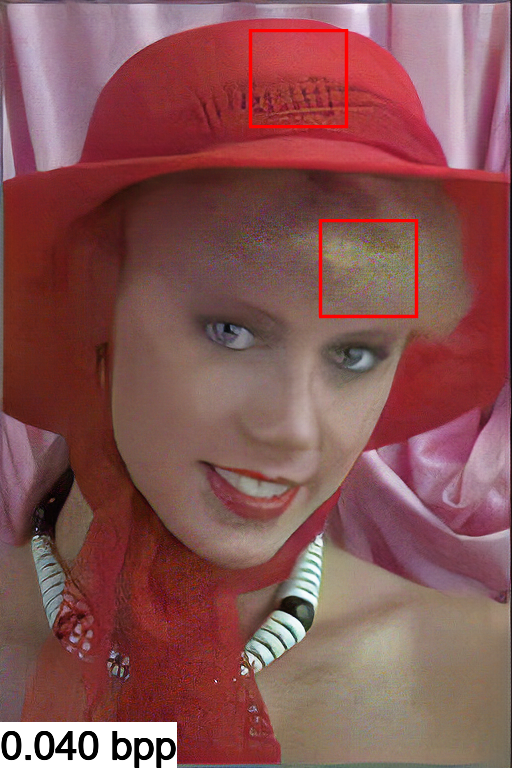}
      \caption{Iwai \etal~\cite{iwai_icpr_2021}}
  \end{subfigure} \hspace*{-0.5em}
  \begin{subfigure}[b]{0.24\textwidth}
    \centering
      \includegraphics[width=\textwidth, interpolate=false]{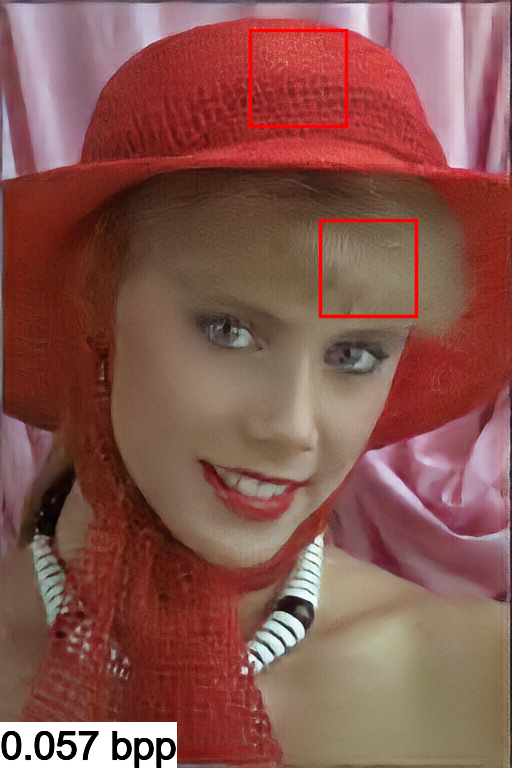}
      \caption{Mentzer \etal~\cite{mentzer_nips_2020}}
  \end{subfigure} \hspace*{-0.5em}
  \begin{subfigure}[b]{0.24\textwidth}
    \centering
      \includegraphics[width=\textwidth, interpolate=false]{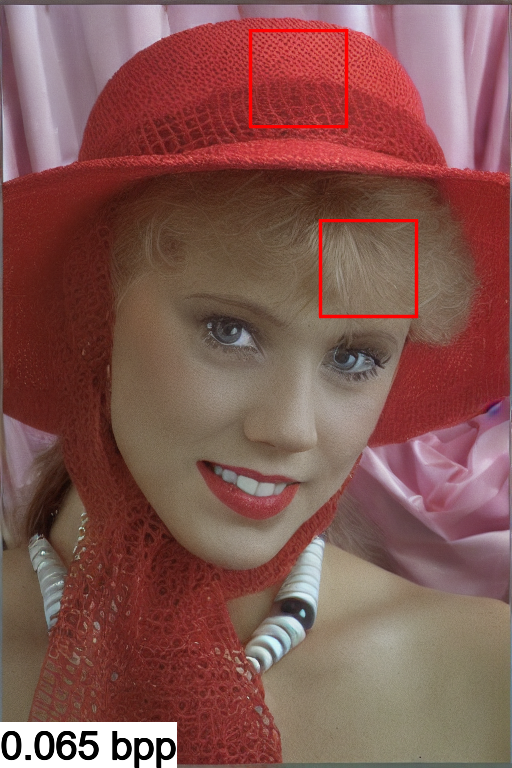}
      \caption{Ours}
  \end{subfigure}
 \end{center}
   \vspace{-9pt}
    \caption{Visual examples of generated images after extreme compression.  Our model has the average bitrate for better performance while two competitive models are subject to severe artifacts due to abnormally low bitrates, a common disadvantage of prior works where models are only trained for a target average bitrate over a large training set.}
   \vspace{-3pt}
\label{fig:k04}
\end{figure*}

For the classifier-free guidance included in the diffusion model, it is generally known that higher guidance scale improves generated image quality.
However, with the introduction of compression guidance in our method, it is useful to validate
the effects of different compression guidance scales by themselves, as well as in combination of different classifier-free guidance scales.
As shown in Fig.~\ref{fig:scales}, we have selected four different image metrics to cover both image reconstruction accuracy and
perceptual quality, where PSNR is used to measure reconstruction accuracy, FSIM and LPIPS are chosen for both accuracy and perceptual quality
and NIQE is for perceptual quality only.  All experiments are conducted on the Kodak dataset.
In general, higher compression guidance scale $s_c$ is preferred for reconstruction accuracy but less favorable for perceptual quality.
While for classifier-free guidance scale $s_f$, it does not have a significant impact on accuracy for the range we tested on while it shows
an optimal value slightly less than 1 for all perceptual quality related metrics.
As a larger value like 5 is often recommended for the classifier-free guidance scale $s_f$ when used for its original generation applications,
this is the first observation that a smaller values less than 1, probably caused by introduction of compression guidance.
For final experiments, $s_c$ and $s_f$ are set empirically as 215 and 0.95 for
the best trade-off between reconstruction accuracy and perceptual quality.

\begin{figure*}[t]
\captionsetup[subfigure]{font=small, labelformat=empty}
\begin{center}
  \begin{subfigure}[b]{0.16\textwidth}
    \centering
      \includegraphics[width=\textwidth, interpolate=false]{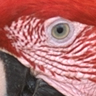}
  \end{subfigure} \hspace*{-0.5em}
  \begin{subfigure}[b]{0.16\textwidth}
    \centering
      \includegraphics[width=\textwidth, interpolate=false]{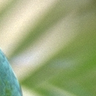}
  \end{subfigure} \hspace*{-0.5em}
  \begin{subfigure}[b]{0.16\textwidth}
    \centering
      \includegraphics[width=\textwidth, interpolate=false]{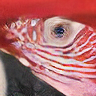}
  \end{subfigure} \hspace*{-0.5em}
  \begin{subfigure}[b]{0.16\textwidth}
    \centering
      \includegraphics[width=\textwidth, interpolate=false]{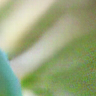}
  \end{subfigure} \hspace*{-0.5em}
  \begin{subfigure}[b]{0.16\textwidth}
    \centering
      \includegraphics[width=\textwidth, interpolate=false]{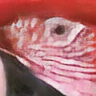}
  \end{subfigure} \hspace*{-0.5em}
  \begin{subfigure}[b]{0.16\textwidth}
    \centering
      \includegraphics[width=\textwidth, interpolate=false]{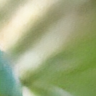}
  \end{subfigure} \hspace*{-0.5em}

  \begin{subfigure}[b]{0.32\textwidth}
    \centering
      \includegraphics[width=\textwidth, interpolate=false]{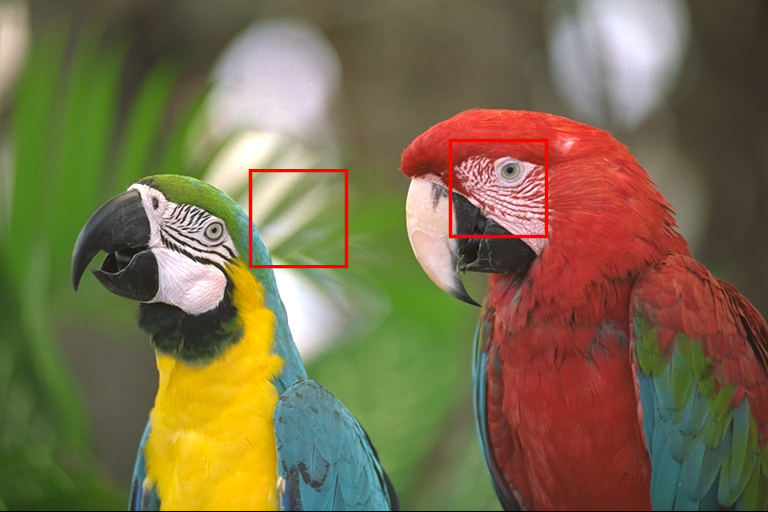}
      \caption{Original}
  \end{subfigure} \hspace*{-0.5em}
  \begin{subfigure}[b]{0.32\textwidth}
    \centering
      \includegraphics[width=\textwidth, interpolate=false]{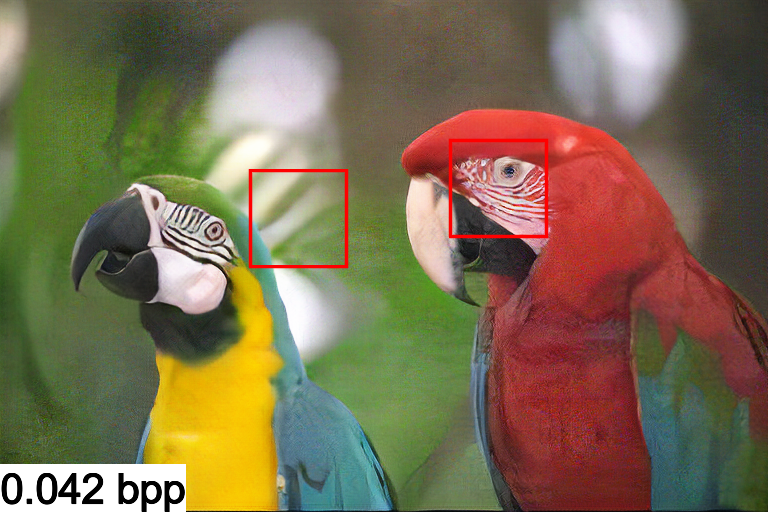}
      \caption{Iwai \etal~\cite{iwai_icpr_2021}}
  \end{subfigure} \hspace*{-0.5em}
  \begin{subfigure}[b]{0.32\textwidth}
    \centering
      \includegraphics[width=\textwidth, interpolate=false]{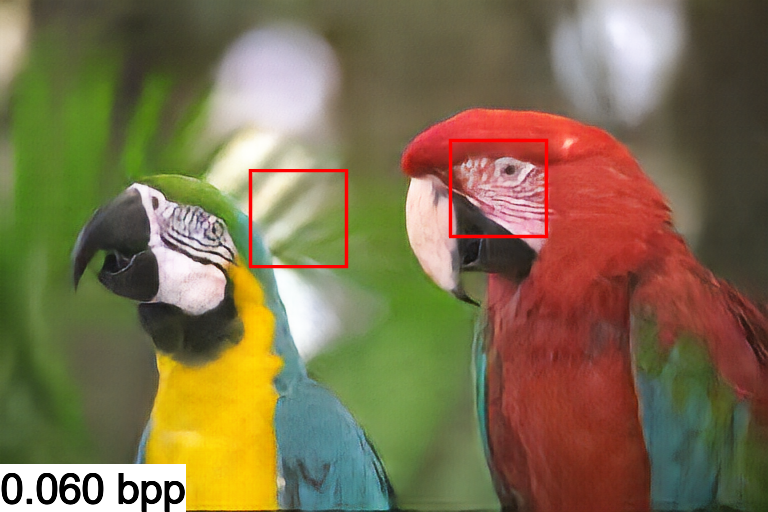}
      \caption{Mentzer \etal~\cite{mentzer_nips_2020}}
  \end{subfigure}
  
  \vspace{6pt}

  \begin{subfigure}[b]{0.16\textwidth}
    \centering
      \includegraphics[width=\textwidth, interpolate=false]{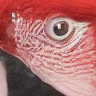}
  \end{subfigure} \hspace*{-0.5em}
  \begin{subfigure}[b]{0.16\textwidth}
    \centering
      \includegraphics[width=\textwidth, interpolate=false]{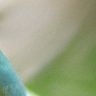}
  \end{subfigure} \hspace*{-0.5em}
  \begin{subfigure}[b]{0.16\textwidth}
    \centering
      \includegraphics[width=\textwidth, interpolate=false]{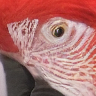}
  \end{subfigure} \hspace*{-0.5em}
  \begin{subfigure}[b]{0.16\textwidth}
    \centering
      \includegraphics[width=\textwidth, interpolate=false]{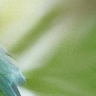}
  \end{subfigure} \hspace*{-0.5em}
  \begin{subfigure}[b]{0.16\textwidth}
    \centering
      \includegraphics[width=\textwidth, interpolate=false]{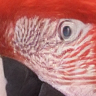}
  \end{subfigure} \hspace*{-0.5em}
  \begin{subfigure}[b]{0.16\textwidth}
    \centering
      \includegraphics[width=\textwidth, interpolate=false]{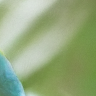}
  \end{subfigure}
  
  \begin{subfigure}[b]{0.32\textwidth}
    \centering
      \includegraphics[width=\textwidth, interpolate=false]{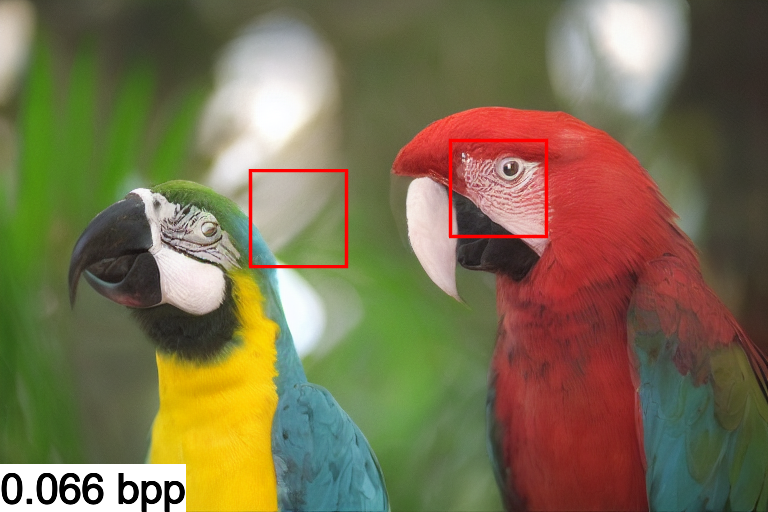}
      \caption{Ours \#1}
  \end{subfigure} \hspace*{-0.5em}
  \begin{subfigure}[b]{0.32\textwidth}
    \centering
      \includegraphics[width=\textwidth, interpolate=false]{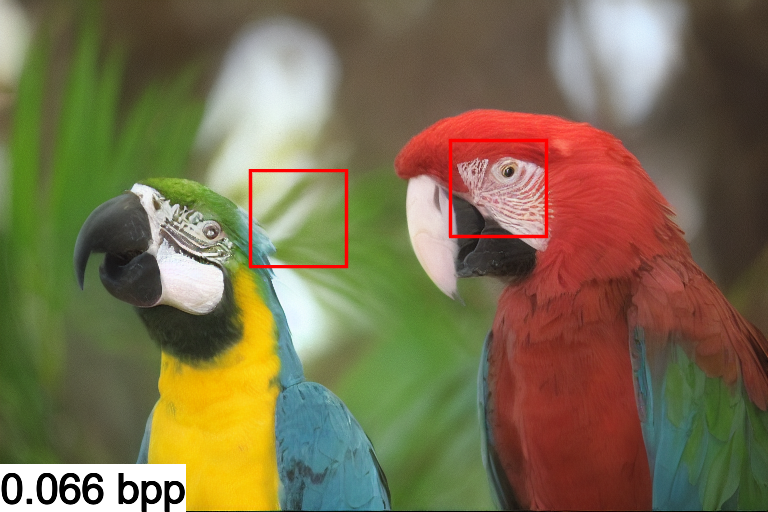}
      \caption{Ours \#2}
  \end{subfigure} \hspace*{-0.5em}
  \begin{subfigure}[b]{0.32\textwidth}
    \centering
      \includegraphics[width=\textwidth, interpolate=false]{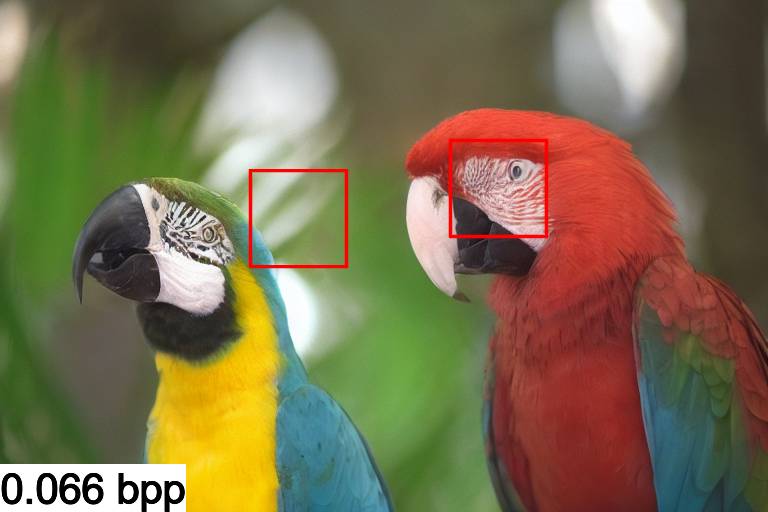}
      \caption{Ours \#1}
  \end{subfigure}
 \end{center}
   \vspace{-9pt}
    \caption{Visual examples of generated images.   Multiple samples from our model using the same compressed source enjoy both high perceptual quality and diversity.}
   \vspace{-3pt}
\label{fig:k23}
\end{figure*}

\subsection{Image Quality Assessment}
We have validated our method in comparison with other state-of-art methods using a comprehensive set of image quality metrics.
For HiFiC developed by Mentzer \etal~\cite{mentzer_nips_2020}, the one with lowest bitrate available is higher than 0.1 bpp.  For fair
comparison, we have retrained the model with a target bitrate of 0.07 bpp using a large set of high quality images, including both
DIV2K and Flickr2K as used in \cite{agustsson_ntire_2017}.  The other one proposed by Iwai \etal~\cite{iwai_icpr_2021} is trained
with a diverse and large dataset COCO~\cite{lin_eccv_2014} so the pre-trained model is used for direct comparison.
For two other methods proposed by Agustsson \etal~\cite{agustsson_iccv_2019} and Dash \etal~\cite{dash_wacv_2020},
the models available are trained the Cityscapes dataset~\cite{cordts_cvpr_2016}, which is limited to city scenes and not sharp enough
comparing to photo-realistic images included in Kodak dataset.  As they are earlier models than the two included for comparison here,
they are not retrained to be included here for assessment.

As shown in Table~\ref{tab:iqa}, our method is the best in perceptual image quality metrics when no ground-truth
reference is available, including NIQE, FID and KID.  For these three
metrics without references, the original uncompressed Kodak images are also assessed to compare with the compressed results.
It shows that the original image indeed have higher perceptual quality compared to compressed ones.
For accuracy related metrics, especially PSNR which is the least related to perceptual quality, our method is not as
impressive as the peers.  In addition to superior perceptual quality, our method has another advantage of near-constant bitrate.
As included in Table~\ref{tab:iqa}, it has a much smaller standard deviation of 0.008 while the other two are 0.021 and 0.028.  
This is useful in applications where transmission bandwidth is very limited and an accurate estimation of bits needed
for storage of image(s) is important before compression.  While the average bitrate for the full Kodak dataset is similar for all
three models, the bitrates for the image shown in Fig.~\ref{fig:k04} are significantly below average for the two model other than
ours.  As a result, generated images from both models are subject to severe blurry artifacts.  In the case of \cite{iwai_icpr_2021}
which has a 0.04 bpp bitrate, it has additional false color artifacts where the full face turns reddish.
In contrast, our method has a sufficient 0.07 bpp bitrate and is able to generate sharp details.

\subsection{Generation Diversity}

While the two competitive models are generative models based on GAN, they are not able to generate high quality images with large
and realistic variations.  In comparison, our model is able to generate photo-realistic images with large diversity in details while
maintaining overall consistency.  As shown by the three sample from our model in Fig.~\ref{fig:k23}, there are large variations
in both foreground and background areas.  For the foreground example, the red line patterns are consistent overall
but vary greatly in details like line sizes and locations, and all three have highly focused sharpness.  In the other example, our samples
all have smooth out-of-focus background yet are quite different from each other.

\section{Conclusions}

In this paper, we present a generative image compression model capable of encoding high resolution image with extremely low bitrates of 0.07 bpp.
It is the first such model built on top of pre-trained text-to-image diffusion models.  Comparing to similar works using GAN, it has some
distinctive advantages: first, it is able to generates diverse images from one compressed source, all with higher perceptual quality and
overall resemblance with the source image;  secondly, it does not need training on a dedicated dataset;
lastly, it has a relatively fixed bitrate for different images, while others models suffer from a large variation in bitrates.

In terms of computational efficiency, both compression and decompression steps of our proposed method are more time consuming in general.
For compression, an iterative learning process is needed to find the optimal compressed textual embedding, while for decompression a large number of denoising steps are needed for high quality outputs.
Beside pure research interest, findings in this study are relevant for some real world applications
where it is not limited by computational resources at both server and client sides but extremely limited in communication bandwidth.

\bibliographystyle{unsrt}

\end{document}